%% file: bare_jrnl_compsoc.tex
\documentclass[10pt,journal,compsoc]{IEEEtran}
\usepackage{ifpdf}

\ifCLASSOPTIONcompsoc
  \usepackage[nocompress]{cite}
\else
  \usepackage{cite}
\fi
\ifCLASSINFOpdf
  \usepackage[pdftex]{graphicx}
  \graphicspath{{figures/}{pictures/}{images/}{./}}
\else
  \usepackage[dvips]{graphicx}
\fi

\usepackage{hyperref}
\usepackage{tabu}                      
\usepackage{booktabs}                  
\usepackage{lipsum}                   
\usepackage{mwe}                       
\usepackage{amsfonts}                   
\usepackage{enumitem}           
\PassOptionsToPackage{warn}{textcomp}  
\usepackage{textcomp}                 
\usepackage{cite}                     
\usepackage{multirow}
\usepackage{makecell}
\usepackage{graphicx}
\usepackage{caption}
\usepackage{xcolor}
\newcommand{\gpt}{{GPT-3}}
\newcommand{\live}{{Live Chart}}
\newcommand{\lives}{{Live Charts}}
\newcommand{\quo}[1]{\textit{``#1''}}
\newcommand{\changed}[1]{{\color{black} {#1}}}
\newcommand{\added}[1]{{#1}}
\newcommand{\minor}[1]{{\color{black} {#1}}}

\begin{document}
\title{Reviving Static Charts into Live Charts}
\author{
    Lu Ying, Yun Wang, Haotian Li, Shuguang Dou,\\ Haidong Zhang, Xinyang Jiang, Huamin Qu, and Yingcai Wu
        \IEEEcompsocitemizethanks{
        \IEEEcompsocthanksitem L. Ying, Y. Wu are with the State Key Lab of CAD\&CG, Zhejiang University, Hangzhou, China. 
        L. Ying is also with Microsoft Research Asia (MSRA), China. 
        E-mail: \{yingluu, ycwu\}@zju.edu.cn. 
        \IEEEcompsocthanksitem  Y. Wang, H. Zhang, X. Jiang is with Microsoft Research Asia, China. 
        E-mail: \{wangyun, haizhang, xinyangjiang\}@microsoft.com.
        \IEEEcompsocthanksitem  H. Li and H. Qu are with The Hong Kong University of Science
        and Technology, Hong Kong, China. 
        E-mail: haotian.li@connect.ust.hk, huamin@cse.ust.hk.
        \IEEEcompsocthanksitem  S. Dou is with the Department of Computer Science and Technology, Tongji University, Shanghai, China.
        He is also with Microsoft Research Asia, Beijing, China. 
        Email: 2010504@tongji.edu.cn.
        \IEEEcompsocthanksitem  Yun Wang is the corresponding author.
        \IEEEcompsocthanksitem This work was done when L. Ying and S. Dou were interns at MSRA.}
    \thanks{Manuscript received xxx xx, 2023; revised xxx xx, 2023.}
}

\markboth{Journal of \LaTeX\ Class Files,~Vol.~x, No.~x, xx~2023}%
{Lu \MakeLowercase{\textit{et al.}}: Reviving Static Charts into Live Charts}

\IEEEtitleabstractindextext{%
\begin{abstract}
    \input{sections/abstract}
\end{abstract}

}

\maketitle

\IEEEdisplaynontitleabstractindextext
\IEEEpeerreviewmaketitle

\input{sections/1_intro.tex}
\input{sections/2_relatedwork}
\input{sections/3_pipeline}

\input{sections/3.1_chart_understanding}
\input{sections/3.2_narration}

\input{sections/3.3_animation}

\input{sections/4.1_use_cases}

\input{sections/4.2_model_performance}
\input{sections/4.3_user_study}
\input{sections/4.4_expert_interview}
\input{sections/5_discussion}
\input{sections/conclusion}

\ifCLASSOPTIONcompsoc
  \section*{Acknowledgments}
  The work was supported by National Key R\&D Program of China (2022YFE0137800), Key ``Pioneer'' R\&D Projects of Zhejiang Province (2023C01120), NSFC (U22A2032), and the Collaborative Innovation Center of Artificial Intelligence by MOE and Zhejiang Provincial Government (ZJU).
  The work was also partially supported by HK RGC GRF grant 16210722.
\else
  \section*{Acknowledgment}
\fi

\ifCLASSOPTIONcaptionsoff
  \newpage
\fi

\bibliographystyle{IEEEtran}
\bibliography{template}

\begin{IEEEbiography}[{\includegraphics[width=1.0in,height=1.25in, clip,keepaspectratio]{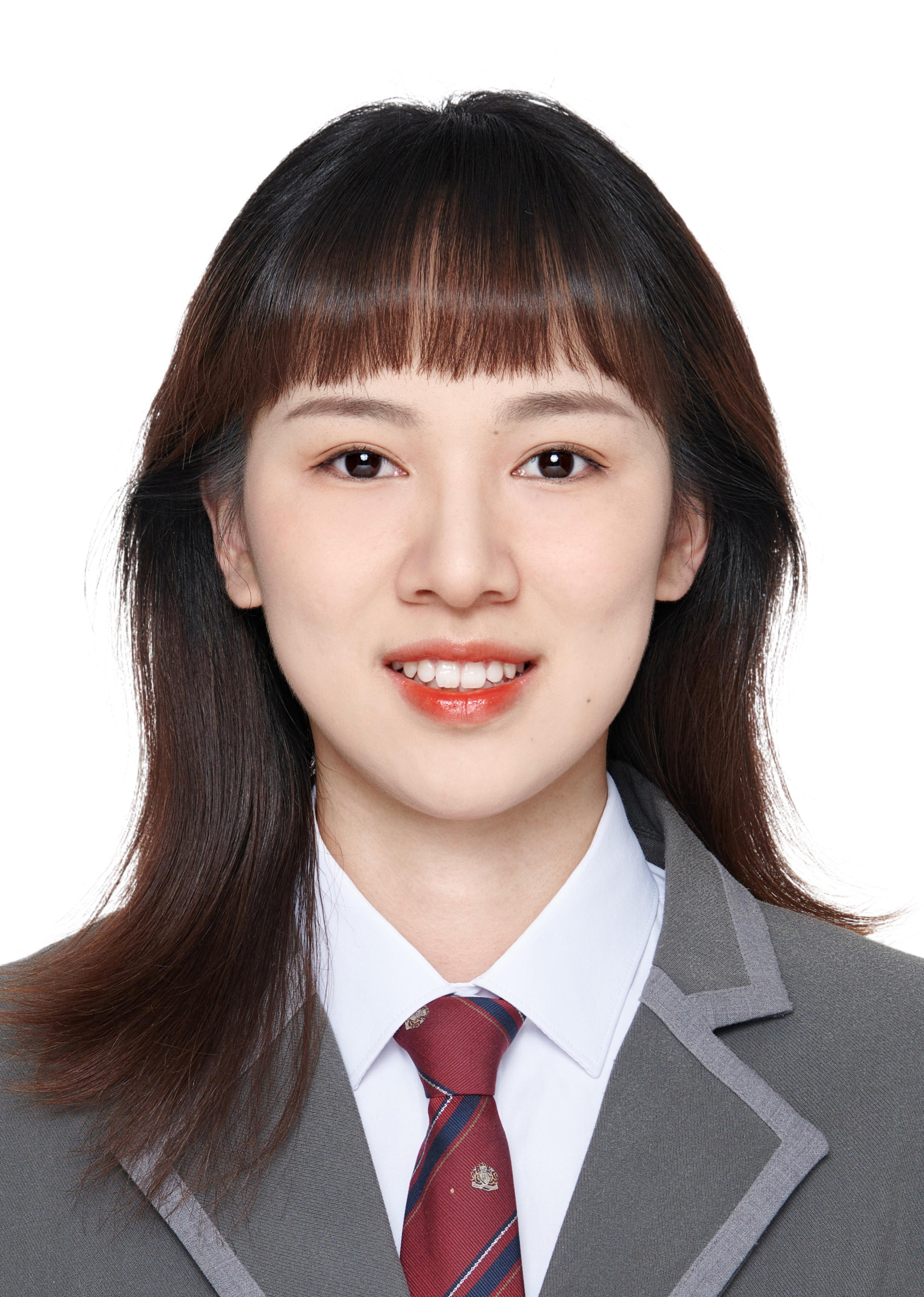}}]{Lu Ying} is currently a Ph.D. candidate at the State Key Lab of CAD\&CG, Zhejiang University. Her main research interests are on data storytelling, glyph-based visualization. She is dedicated to integrating the AI technique into visualization to ease the creation of visualization. She received her BEng in Digital Media Technology from Zhejiang University. For more details, please refer to \url{https://yiyinyinguu.github.io/}.
\end{IEEEbiography}

\begin{IEEEbiography}
[{\includegraphics[width=1in,height=1.25in, clip,keepaspectratio]{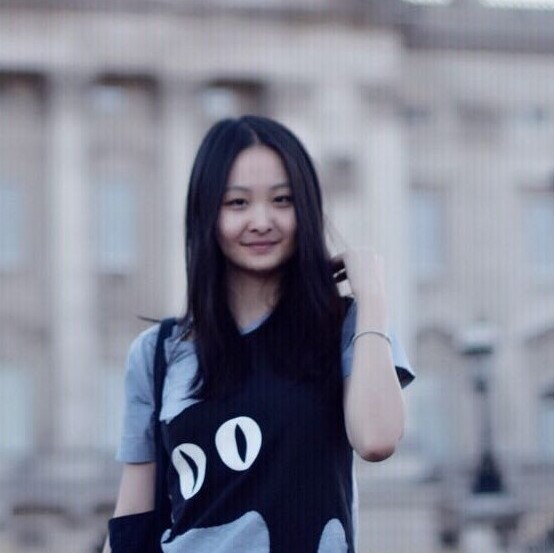}}]{Yun Wang} is a senior researcher in the Data, Knowledge, Intelligence (DKI) Area at Microsoft Research Asia. Her research lies in the intersection of Human-Computer Interaction (HCI), Information Visualization (VIS), Artificial Intelligence (AI), and Data Science (DS). Her work facilitates Human-Data Interaction, Human-AI Collaboration, and Data Storytelling through an interdisciplinary approach. For more details, please refer to \url{https://www.microsoft.com/en-us/research/people/wangyun/}.
\end{IEEEbiography}

\begin{IEEEbiography}[{\includegraphics[width=1.0in,height=1.25in, clip,keepaspectratio]{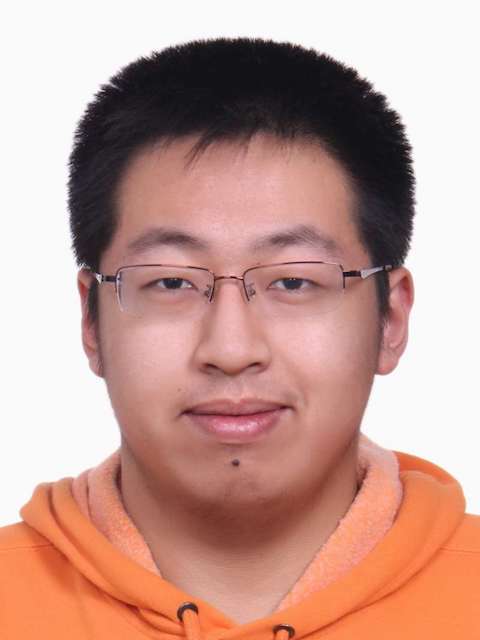}}]{Haotian Li} is currently a Ph.D. candidate in Computer Science and Engineering at the Hong Kong University of Science and Technology (HKUST). His main research interests are data visualization, visual analytics, human-computer interaction and online education. He received his BEng in Computer Engineering from HKUST. For more details, please refer to \url{https://haotian-li.com/}.
\end{IEEEbiography}

\begin{IEEEbiography}[{\includegraphics[width=1in,height=1.25in,clip,keepaspectratio]{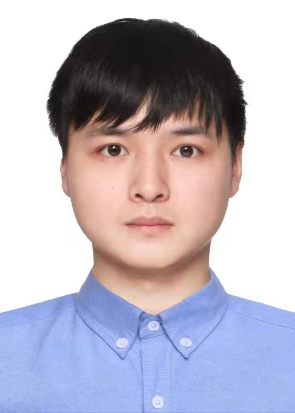}}]{Shuguang Dou}
is currently pursuing a Ph.D. degree with the College of Electronics and Information Engineering, at Tongji University, Shanghai, China. His research interests include Infographic understanding, person re-identification, NAS benchmark, X-ray, and remote sensing. He has published papers in ICLR, AAAI, TIP, TIFS, and TCSVT. He is a program member of ICML, NIPS, and ICLR. For more information, please visit \url{https://shuguang-52.github.io/}.
\end{IEEEbiography}

\begin{IEEEbiography}[{\includegraphics[width=1in,height=1.25in,clip,keepaspectratio]{figures/bio/jiang_xinyang.jpg}}]{Xinyang Jiang} is now a researcher from Microsoft Research Asia. Before joining MSRA, he was a researcher from Tencent Youtu Lab. He received B.E. from Zhejiang University in 2012 and a Ph.D. from Zhejiang University in 2017. His main research areas include cross-modal retrieval, computer vision, and pedestrian re-identification. He has published more than ten papers in CVPR, ICML, NIPS, ICLR,  ECCV, AAAI, ACMMM, TIP and other top conferences and journals on computer vision and artificial intelligence. He is a program member of AAAI, CVPR, MM and other conferences, and a reviewer for TCSVT, TIP and other journals. 
\end{IEEEbiography}

\begin{IEEEbiography}[{\includegraphics[width=1in,height=1.25in,clip,keepaspectratio]{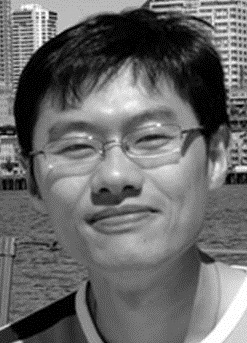}}]{Haidong Zhang} is currently a principal architect with Microsoft Research Asia. He received a Ph.D. degree in computer science from Peking University, China. His research interests include visualization and human-computer interaction.
\end{IEEEbiography}

\begin{IEEEbiography}[{\includegraphics[width=1.0in,height=1.25in, clip,keepaspectratio]{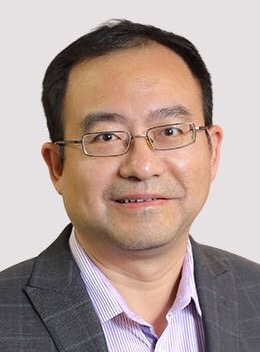}}]{Huamin Qu} is a chair professor in the Department of Computer Science and Engineering (CSE) at the Hong Kong University of Science and Technology (HKUST) and also the director of the interdisciplinary program office (IPO) of HKUST. He obtained a BS in Mathematics from Xi'an Jiaotong University, China, an MS and a PhD in Computer Science from the Stony Brook University. His main research interests are in visualization and human-computer interaction, with focuses on urban informatics, social network analysis, E-learning, text visualization, and explainable artificial intelligence (XAI). For more information, please visit \url{http://huamin.org/}.
\end{IEEEbiography}

\begin{IEEEbiography}[{\includegraphics[width=1in,height=1.25in,clip,keepaspectratio]{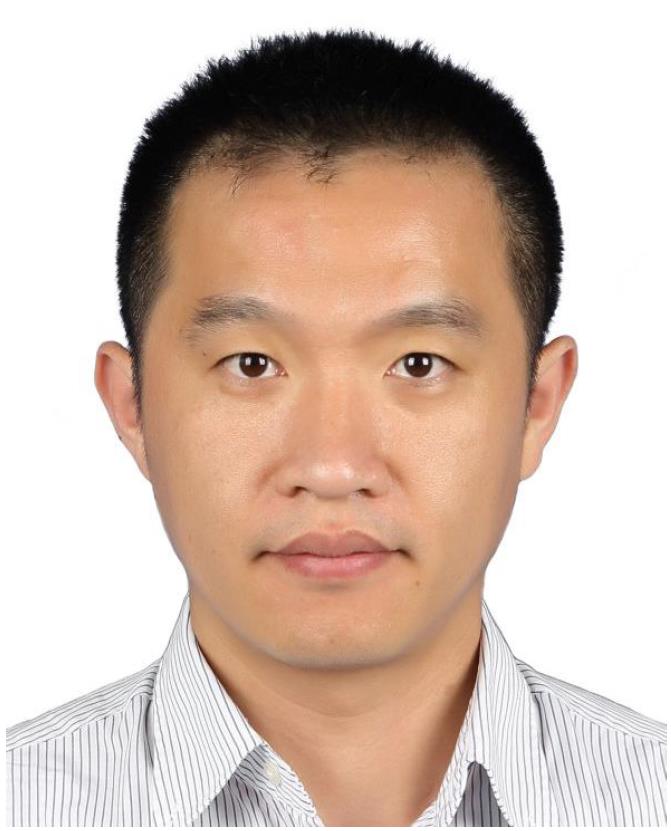}}]{Dr. Yingcai Wu} is a Professor at the State Key Lab of CAD\&CG, Zhejiang University.
His main research interests are information visualization and visual analytics, with focuses on urban computing, sports science, immersive visualization, and social media analysis. 
He received his Ph.D. degree in Computer Science from The Hong Kong University of Science and Technology. 
Prior to his current position, Dr. Wu was a postdoctoral researcher in the University of California, Davis from 2010 to 2012, and a researcher in Microsoft Research Asia from 2012 to 2015. 
For more information, please visit \url{http://www.ycwu.org}.
\end{IEEEbiography}

\end{document}

%% file: sections/abstract.tex
Data charts are prevalent across various fields due to their efficacy in conveying complex data relationships. However, static charts may sometimes struggle to engage readers and efficiently present intricate information, potentially resulting in limited understanding. We introduce ``Live Charts,'' a new format of presentation that decomposes complex information within a chart and explains the information pieces sequentially through rich animations and accompanying audio narration. 
We propose an automated approach to revive static charts into Live Charts. Our method integrates GNN-based techniques to analyze the chart components and extract data from charts. Then we adopt large natural language models to generate appropriate animated visuals along with a voice-over to produce Live Charts from static ones. 
We conducted a thorough evaluation of our approach, which involved the model performance, use cases, a crowd-sourced user study, and expert interviews. The results demonstrate Live Charts offer a multi-sensory experience where readers can follow the information and understand the data insights better. We analyze the benefits and drawbacks of Live Charts over static charts as a new information consumption experience.

%% file: sections/1_intro.tex
\section{Introduction}

\IEEEPARstart{S}{tatistical} charts are widely used to deliver visually appealing and informative presentations of data.
They have been used in various fields, such as urban~\cite{zhao2022uncertainty}, education~\cite{zhang2022better}, and science~\cite{li2023causality}, to communicate data patterns~\cite{shirato2023identifying} and insights. 

However, static charts have limitations due to their unchanging nature. 
The absence of a clear reading order can confuse as all information is presented simultaneously, and users do not know where to start or how to locate the corresponding elements~\cite{amini2018hooked}.
Further, static charts have a limited capacity for conveying information~\cite{ren2017chartaccent}.
A chart containing excessive visual components or encodings may overwhelm users, leading to a visual burden. 
Moreover, static charts may not be engaging enough to maintain users' interest~\cite{chevalier2016animations}, reducing their effectiveness as a communication tool.
These phenomena emphasize the need for strategies to enhance data communication via charts.

There have been several studies aimed at enhancing static charts with various solutions. 
One approach is to break down the complex original chart information and introduce pieces of information gradually, making it easier for users to comprehend~\cite{deng2022revisiting}. 
For instance, recent research~\cite{wang2018narvis, wang2021animated} hierarchically decomposes a visualization or infographic design and introduces the compositions progressively.
Alternatively, some studies have attempted to augment static charts' expressiveness by including available information, such as adding question-answering~\cite{kahou2018figureqa}, annotations \cite{lai2020automatic} or captions \cite{mahmood2014automated, balajicharttexta}.

\begin{figure*}[htbp]
    \centering
    \includegraphics{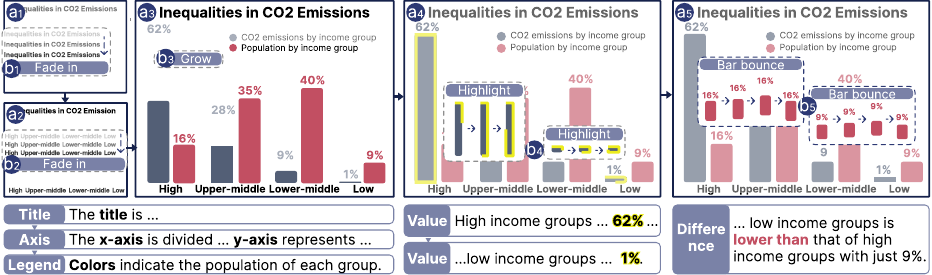}
    \caption{An example of a Live Chart.
        (a1-a5) The sequential process of the Live Chart.
        (b1-b5) The animations in the chart.
        The texts below are the audio narration for the corresponding frames, with the first tag indicating the chart component and the insight type.}
    \label{fig:teaser}
\end{figure*}

In this paper, we introduce Live Charts, a new format that revitalizes static charts by integrating the advantages of simplifying complex information and incorporating explanatory details. 
Our approach employs a multi-sensory technique, including audio as an additional source of information for users. 
Live Charts present data sequentially, accompanied by audio narration that introduces visual encodings and elucidates the messages conveyed by the charts. 
For instance, when examining a static bar chart such as the one depicted in Fig.~\ref{fig:teaser}(a3), a Live Chart presentation would involve increasing the height of the red bars synchronously while narrating stockpile data about the United States. 
Adopting this format for data visualization, we explore how Live Charts may enable users to comprehend the presented information more easily and accurately.

Nevertheless, if manual creation is required for Live Charts, this new format of presentation may be deemed less useful due to the numerous steps involved, such as manually extracting data and visual encodings, designing animations, and crafting a chart narration that synchronizes with the animation. 
To address this concern, we further explore methods to automate the creation of Live Charts from static ones.
We break down the procedure into a series of computational methods that can be automated, thereby enhancing the efficiency of this new format.
\changed{We trained a Graph Neural Network (GNN) model to recover the embedded data and visual encodings of static charts on our own generated chart dataset.}
Subsequently, we aim to create the audio narration.
Given that data insights are frequently used to create narratives~\cite{srinivasan2019augmenting}, we expect our narration to contain a compelling data story showcased together with the Live Chart.
We employ large language models (LLMs) to uncover intriguing insights within the data and then generate the corresponding narration.
Ultimately, we integrate the narration with the static chart to create a Live Chart by animating relevant visual elements based on the extracted insights and corresponding narration.

We evaluate the effectiveness of our approach in different aspects. We first present a variety of real-world use cases that highlight the practical applications of our methods. 
Following that, we showcase the performance of our chart understanding model, which serves as the foundation for our automated technique.
Furthermore, we conduct a crowd-sourced user study to investigate our generated Live Charts and conduct expert interviews to further evaluate our automated approach.
Finally, we discuss the insights gained from our research and propose potential avenues for future exploration.

Our primary contribution is an automatic pipeline for reviving static charts into Live Charts.
The approach consists of 1) a method including a dual-stream GNN that recovers data in SVG charts and 2) an automatic enhancement from recovered data by incorporating narrations and animations, resulting in the generation of Live Charts.
To validate the efficacy of our approach, we conducted comprehensive evaluations.
The results provide evidence supporting the effectiveness of our chart revival approach and the resulting Live Chart format.

%% file: sections/2_relatedwork.tex
\section{Related Work}

In this section, we review research on visualization understanding, visualization enhancement and data video generation.
\subsection{Visualization Understanding}
The proliferation of visualizations gives rise to research efforts to understand existing images automatically~\cite{davila2021charta}.
Given visualization images, researchers seek to understand their content (e.g., chart type, data, and visual encodings).
Based on the image input, prior works can be mainly categorized into two classes: raster image and vector image.

Researchers have explored the interpretation of raster images from different perspectives~\cite{gao2012view, poco2017reverseengineering}.
For instance, Savva et al.~\cite{savva2011revision} proposed ReVision, which identified chart types and extracted data from bitmap charts.
Later Davila et al.~\cite{davila2021charta} conducted a survey that covered the topic of automated data extraction from chart images.
Despite the support of extracting various information from chart images, raster images are hard to reuse or add extra graphic elements onto the image~\cite{masson2023chartdetective}.
Harper and Agrawala~\cite{harper2014deconstructing, harper2017converting} focused on vector images and used D3 chart~\cite{bostock2011DataDriven} as the input to extract the complete chart structure.
With the embedded information, the charts can be utilized in multiple scenes, such as restyle~\cite{harper2014deconstructing}, template extracting~\cite{harper2017converting}, and style searching~\cite{hoque2019searchinga}.
Kim et al.~\cite{kim2020answering} focused on Vega-Lite charts~\cite{satyanarayanvegalite} and extracted data and visual encodings in two stages.
Recently, Masson et al.~\cite{masson2023chartdetective} proposed an interactive method to extract data from vector charts.

Our approach also takes vector charts as the input since we need to motion the image.
We use charts in SVG format with structural information~\cite{li2022structure}.
Moreover, the structures of charts generated by different tools are distinct (e.g., D3, Vega-lite, and Adobe Illustrator).
Compared with the methods that targeted a specific SVG type, our model's structure is not influenced by the format of SVG files. 
Specially, we analyze the structure information for online SVGs automatically by GNN techniques, including the data and visual encodings.

\subsection{Visualization Enhancement}
Visualization enhancement has attracted researchers' interest in recent years with the evolving techniques of visual understanding~\cite{kai2023comantics}, which may assist users in understanding the visualization quickly and effectively.
The methods can be mainly categorized into three categories, question-answering, annotation, and caption.

Question answering refers to generating answers given a question. 
Some studies include DQVA~\cite{kafle2018dvqaa}, FigureQA~\cite{kahou2018figureqa}, PlotQA~\cite{methani2020plotqa} contributed datasets and models for different plots.
Further, researchers explored new methods (e.g., LEAF-QA~\cite{chaudhry2020leafqa}, FigureNet~\cite{reddy2019figurenet}) to achieve higher accuracy in question-answering tasks.
Compared with treating the input chart image using computer vision techniques, Kim et al.~\cite{kim2020answering} focused on Vega-lite charts and extracted the chart structure to answer natural language questions.

Annotation, as the extra graphical or textual elements~\cite{munzner2014visualization}, plays a vital role in conveying information in data-driven storytelling~\cite{ren2017chartaccent} and attracting users' attention to specific parts of the visualization~\cite{bongshinlee2013sketchstory}.
Kong and Agrawala~\cite{kong2012graphicala} presented an automatic system to aid chart reading by generating desired overlays.
Bryan et al.~\cite{bryan2017temporal} proposed Temporal Summary Images to assist visualization analysis via interactive annotation by inputting images and raw data.
Lai et al.~\cite{lai2020automatic} adopted textual descriptions and images as the input.
They introduced an automatic pipeline for annotating visualization.
Unlike the raster images in the above work, Lu et al.~\cite{minlu2017interaction} focused on web-based visualizations and developed a system to augment them via a palette of interactions.
Ren et al.~\cite{ren2017chartaccent} characterized a design space of annotation and implemented ChartAccent, which enables users to easily enhance SVG-based charts by a series of annotation interactions.

The caption provides a summary for a visualization.
Mittal et al.~\cite{mittal1998describing} presented a system, which uses a text planner to select the caption content and structure.
Recently, researchers utilized template-based descriptions to depict charts in certain genres of narrative visualization~\cite{shi2021calliope}, such as slideshows~\cite{li2023notable}.
With the development of computer vision and natural language processing techniques, researchers contributed plenty of models to generate accurate descriptions given rasterized charts~\cite{kantharaj2022charttotext, obeid2020charttotext}.
Liu et al.~\cite{liu2020autocaption} targeted vector images and proposed an approach including several models to generate captions automatically.

Our work aligns with the direction to augment the chart but goes beyond to revive the charts into Live Charts.

\subsection{Data Video Generation}
Plenty of research is dedicated to enhancing data visualization through creating data videos~\cite{amini2015understanding}. 
Many studies focus on integrating animations and developing systems to generate animations from data~\cite{amini2017authoring, shi2021autoclips, shin2022roslingifier}. 
For example, DataClips~\cite{amini2017authoring} allows non-experts to generate data videos by inputting data. 
Through a series of interactions, users can select and integrate data-driven clips from the given library and form a video.
DataParticles~\cite{cao2023dataparticles} allows creators to author animated unit visualization in data stories.
AutoClips~\cite{shi2021autoclips} moves forward to save human effort by automatically crafting data videos from a sequence of data facts.
However, the above work cannot revive static charts with image input.
There are alternative approaches that begin with specific image inputs.
For instance, Data Animator~\cite{thompson2021data} chooses the Data Illustrator~\cite{liu2018data} file format as the input, which includes detailed data and visual encoding information and enables authors to generate animated data graphics and transitions without programming.
Based on Canis~\cite{ge2020canis}, Ge et al.~\cite{ge2021cast} designed CAST, an interactive authoring tool to create data videos by inputting data-enriched SVGs.
However, these methods have specific input requirements and cannot process SVGs without data.

Though these tools assist users in generating animations, much effort is still required to transform static charts into Live Charts. 
Focused on infographics, Infomotion~\cite{wang2021animated} allows users to automatically generate animated presentations from static infographics.
However, it does not assist basic charts. Moreover, current studies tend to neglect the importance of audio narration, which is a crucial aspect of creating a comprehensive data video. 
As a result, users must still use additional tools to record audio narration and synchronize it with animation. 
In our research, we propose methods for automatically deriving Live Charts from static ones and consider both the animation and audio narration simultaneously in the process.

%% file: sections/3_pipeline.tex
\section{Live Charts}
\label{sec: livecharts}
Since making Live Charts from static vector-based charts requires considerable effort, we aim to revive them into Live Charts automatically.

To enhance the chart expression, previous studies have emphasized the significance of annotations or textual cues to enhance user comprehension of charts~\cite{ren2017chartaccent}.
In contrast to traditional text cues near charts that divert visual attention, we utilize voiceover narration—freeing up visual focus while enhancing chart comprehension.
Moreover, prior research~\cite{wang2018narvis} highlights the effectiveness of breaking down complex information into smaller, more digestible chunks when presenting charts.
In this regard, we choose to integrate animations into Live Charts, which can assist users in visually processing intricate data and comprehending complex chart representations.
Furthermore, a multi-sensory experience by synchronizing narrations and animations within the Live Charts can complement each other by providing supplementary details, or indirectly link to each other by presenting data from different perspectives~\cite{cheng2022investigating}.
Therefore, we propose an approach that involves three steps:
\begin{enumerate}[noitemsep]
    \item[(1)] Using GNN-based techniques, we identify the data together with the visual encodings from the input SVG.
    \item[(2)] We derive the audio narration with insights and its corresponding data with LLMs.
    \item[(3)] For some expressions in narration, we design the corresponding animations and synchronize them to the audio.
\end{enumerate}

\begin{figure}[htbp]
    \centering
    \includegraphics{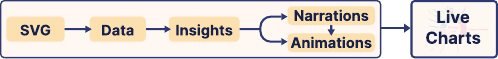}
    \caption{Our automatic approach for generating Live Charts.}
    \label{fig:pipeline}
\end{figure}

\changed{
We then describe each step in detail accompanied by an example depicted in \autoref{fig:ChartUnder}(a), which we refer to as the ``airport chart.''
}

%% file: sections/3.1_chart_understanding.tex
\subsection{Chart Understanding}
\label{sec: understanding}

\begin{figure*}[htbp]
\begin{center}
\includegraphics{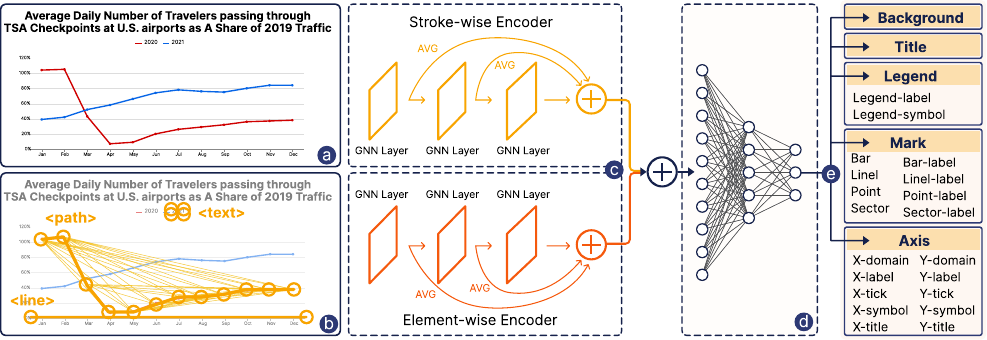}
\end{center}
\caption{A GNN-based approach to classify chart elements for a chart. 
(a) Original SVG-based chart; 
(b) Conversion from SVG to graph; 
(c) Two different vector encoders to extract features; 
(d) Classification of each element by a multi-layer perceptron. 
(e) Categories for elements in the chart. Each box corresponds to a primary category, followed by a set of sub-categories.}
\label{fig:ChartUnder}
\end{figure*}

In this section, we present an approach for automatically understanding charts and recovering data together with` visual encodings. 

Compared to common existing methods that rely on pixel-based charts, our proposed approach directly handles the raw SVG files to revive charts into Live Charts. 
Pixel-based methods often suffer from high computational overheads and limitations in modifying individual elements due to their rasterized nature~\cite{yolat, rendnet}. 
In contrast, SVG files, being vector-based, offer efficient rendering regardless of resolution and allow for targeted modifications through the use of element properties.
The structured information between elements in vector graphics allows us to model SVG elements, such as lines, paths, and circles, as a multi-graph, which effectively captures the essential structural and spatial information. 
In light of the proficiency of GNNs in handling structured graph-based data and their success in previous research~\cite{li2022structure}, we leverage GNNs for recognizing and analyzing elements within SVG files. 

Especially, our chart understanding method involves a two-stage process: recognizing chart elements (e.g., marks, legends) and recovering visual encodings. 
First, we apply a GNN-based method~\cite{yolat} to detect chart elements and classify them into pre-defined categories. 
Second, we establish a set of rules to extract data and recover visual encodings based on the results from the classification.

\subsubsection{Chart Element Recognition}
\label{sec: recognition}
Our goal is to identify various chart elements within a given chart. 
To utilize the structural information within SVGs, we first convert the chart to graphs and apply a GNN-based method to extract features with two vector encoders.
Then, we predict each element's category based on the extracted features.

\textbf{Convert Chart to Graph.}
We begin by constructing nodes and then connecting them with edges.
SVG-based charts are represented as a set of SVG elements, including path, text, line, and other basic elements. 
Two types of SVG elements (graphic and textual) are handled separately, as they possess distinct attributes.
For graphic elements such as paths and lines, we utilize each element's start and end points as two nodes of the corresponding graph.
For text elements, we connect the vertices of the bounding box to form a 4-node graph.
Yellow circles in \autoref{fig:ChartUnder}(b) provide examples of the nodes.
Next, a feature vector consisting of five dimensions was created to represent each node, including the element type ($e_t$), which is converted into a one-hot vector, the spatial coordinates of the points ($pos$), the fill color ($e_f$), the stroke color ($e_{sc}$), and the stroke width ($e_{sw}$).
Thus, the node feature $V_f$ is:
\begin{equation}
    V_f = \{e_t, pos, e_{f}, e_{sc}, e_{sw}\}.
\end{equation}

After constructing nodes, we proceed to connect them using two types of edges.
The first type is stroke-wise edges ($E_s$), which connect nodes that are directly connected in the SVG, such as adjacent points in a line chart.
These edges are depicted by thick yellow lines in \autoref{fig:ChartUnder}(b).

\begin{equation}
    E_s = \{(v_i, v_j)|v_i, v_j \in \mathbb{C} \},
\end{equation}
where $\mathbb{C}$ is a set of tuples containing the start point $v_i$ and end point $v_j$.

The second type, element-wise edges ($E_e$), connect nodes that have the same element type $e_t$.
These connections can be observed as thin yellow lines in \autoref{fig:ChartUnder}(b):

\begin{equation}
    E_e = \{(v_i, v_j)|v_i, v_j\in e_i\}, i\in N.
\end{equation}

After that, we obtain the graph $G(V, E_s, E_e)$ with several nodes and the connected edges, where $V$ denotes the node, $E_s$ denotes the stroke-wise edge, and $E_e$ denotes the element-wise edge.

\textbf{Feature Extraction with two Vector Encoders.}
After converting the chart into a graph, we extract features from it using two vector encoders, each tailored for the corresponding category of edges. 
For stroke-wise edges, we utilize a stroke-wise encoder, using a GNN layer that incorporates the functions of two linear transformations. 
For element-wise edges, given the larger number of edges, we employ a GNN layer with one linear transformation. The specific architecture of the two encoders is depicted in \autoref{fig:ChartUnder}(c), where the features derived from each layer of the encoders are aggregated and utilized for downstream tasks.
\begin{equation}
    F_{fuse} = concat(f^1_s, f^2_s, \cdots, f^l_s, f^1_e, f^2_e, \cdots, f^l_e),
\end{equation}
where $F_{fuse}$ is the fused features of a specific element.

\textbf{Classification Task.} 
For each SVG element, we obtain the graph representation by aggregating the features extracted by two vector encoders. 
To comprehend the categories that are depicted by each element, we utilize a multi-layer perceptron to classify the graph representation that corresponds to each element.
We define five primary categories: background, title, legend, mark, and axis. 
The corresponding sub-categories can be found in \autoref{fig:ChartUnder}(e).
The classification task is performed across these class labels, with cross-entropy losses serving as the training objective for the entire model.

\textbf{Dataset.}
To train the above model, we require an SVG chart dataset that encompasses different types of charts. 
Additionally, the ground truth information for each element is essential for indicating its corresponding category in the chart. 
We decided to use Vega-Lite~\cite{satyanarayanvegalite}, Plotly~\footnote{https://plotly.com/} and D3~\cite{bostock2011d3} for generating charts instead of collecting due to the substantial amount of data required.
The high-level grammar enables the creation of charts in various types and styles while striking a balance between customization and complexity~\cite{satyanarayanvegalite}.
We utilized the JavaScript library ``Faker''\footnote{\added{https://fakerjs.dev/}} to populate the charts with synthetic data.
In this paper, we consider three basic chart types: bar chart, line chart, and pie (donut) chart.
We conducted a comprehensive analysis of different chart components, such as background, axis, and points, based on previous collections~\cite{ren2017chartaccent}. We aim to identify the diverse styles associated with each component. Following that, we made a comprehensive list of styles, which can be used for generating charts.
Consequently, for Vega-lite, Plotly, and D3, we generated 3,000 charts for each chart type separately, resulting in a total of 9,000 charts for each dataset~\footnote{https://github.com/YiYinYinguu/SVG-Chart-Dataset}.
The performance of our model on this dataset is evaluated in \autoref{sec: performance}.

\subsubsection{Data and Visual Encoding Recovering}
We aim to recover the data and visual encodings for the above categorized elements.
This is achieved by analyzing the relationships between the classified elements in the SVG. 
We then add the data and visual encodings into the original SVG and produce a labeled SVG to facilitate the next two steps for generating the Live Chart. 
The process contains three steps:

\begin{figure}[htbp]
    \centering
    \includegraphics{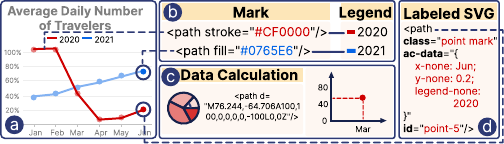}
    \caption{The process of recovering visual encodings. (a) The example chart. Dotted dark blue lines connect chart elements with their SVG expressions. (b) Matching the mark with the legend. (c) Data calculation for different charts. (d) One path in the labeled SVG. }
    \label{fig:visual_encoding}
\end{figure}

\textbf{Legend-Mark Matching.}
With the aid of the classification results, we can obtain the bounding box of the legend labels and legend symbols through the attribute $d$ in the path element. 
For charts that include legends, we determine the corresponding label for each legend symbol by calculating the Euclidean distance between the top-left corner of the bounding box of each symbol and label and identifying the nearest label.
Next, we match each mark with the corresponding legend by comparing the stroke and fill color of the mark and legend symbols (\autoref{fig:visual_encoding}(b)).

\textbf{Data Calculation.}
For charts that contain textual information, such as bar labels appearing alongside the bars, we match each mark with the closest label to retrieve the corresponding y-value.
For charts that lack textual information, we handle the Polar (pie charts) and Cartesian (line and bar charts) coordinate systems separately, as shown in \autoref{fig:visual_encoding}(c).
For the Polar coordinate system, we parse the $d$ attribute for the path element and get the arc length of all sectors. We then calculate the percentage of data represented by each sector.
For the Cartesian coordinate system, the marks are first matched with the closest ``x-label'' in horizontal distance to retrieve the corresponding x-value.
Then, the y-value is calculated using an interpolation method with all values of the ``y-label'' and the vertical distance between the marks and the closest``y-label''.

\textbf{Output Labeled SVG.}
After we calculate the data within the chart, we add it to the original SVG.
The output labeled SVG $C_{lsvg}$ contains the same structure as the original SVG, but with additional attributes that represent the extra information extracted from the chart.
\autoref{fig:visual_encoding}(d) shows an example <path> in the labeled SVG.
We add three attributes:
\begin{itemize}
\item\textbf{Class}: We assign two class names, including the primary category and the sub-category type (\autoref{fig:ChartUnder}(e)), to the element.
\item\textbf{ID}: We define the ID as the concatenation of the sub-category type and an ordered number (e.g., \textit{line-0}). 
This attribute is used for identifying specific elements within the SVG in the following workflow.
\item\textbf{Ac-data}: 
We store the extracted data in this attribute, using two data dimensions for charts without legends and three dimensions for charts with legends. 
We assign the content of x-title, y-title, and legend-title as the field names for these dimensions. 
In cases where these fields are empty, such as the x, y, legend fields in the line chart depicted in \autoref{fig:visual_encoding}(a), we label them as ``None'' (\autoref{fig:visual_encoding}(d)).
\end{itemize}

%% file: sections/3.2_narration.tex
\subsection{Narration}
\label{sec:narration}

\begin{figure*}[htbp]
\centering
\includegraphics{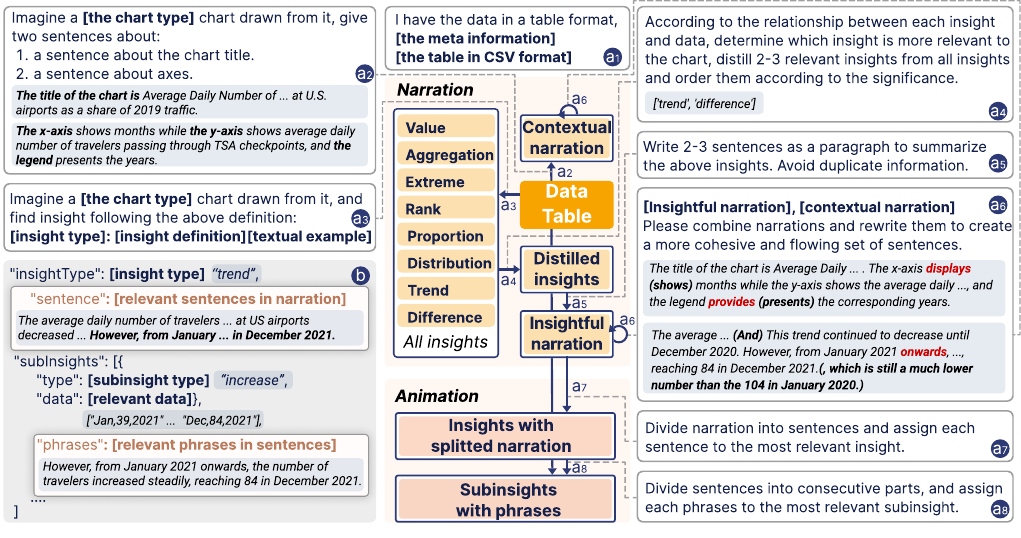}
\caption{
\changed{Creating narration and animations involves multiple sub-tasks for \gpt{}, depicted by the inner blue boxes and arrows.
The starting position of the arrow indicates the information contained within the prompt (with ``Data Table'' being a necessary element for each prompt). 
Each blue box represents the output generated by \gpt{}.
(a1)-(a8) are part of prompt components.
We use dotted gray lines and numbers to illustrate the prompt usage during the process.
(b) The insight structure in JSON format.
Sentences and phrases will be added in animation generation.
Italic text with a blue background represents an example of the prompt output, using the ``airport chart'' in \autoref{fig:ChartUnder}(a), to demonstrate how the process works.
}} 
\label{fig:narration}
\end{figure*}

In this section, we introduce methods to generate audio narration.
High-quality audio narrations can add informative context and direct users' attention~\cite{rubab2023exploring}.
From the previous analysis~\cite{cheng2022investigating}, narration often involves data contexts and insights.
We accept a similar structure: contextual information following interesting insights.

Researchers have adopted various methods to generate insights~\cite{ma2021MetaInsight, ding2019QuickInsights, deng2022dashbot}.
Inspired by the remarkable proficiency of large language models (LLMs) in handling open-ended queries, we select \gpt{} to generate narrations.
When presented with a text $prompt$, the \gpt{} model generates a string-format answer. 
Adhering to the fundamental principles~\footnote{https://beta.openai.com/docs/guides/completion/prompt-design} of ``show and tell'', ``provide quality data'', and ``check settings'', we devise distinct prompts for the following various tasks.

Since GPT-3 cannot process images directly, we use the chart specification as input. 
We create a data table by extracting the ``ac-data'' attribute from the labeled SVG.
In addition, we gather meta information such as the chart's title, x-axis, and y-axis labels. 
This data table and meta-information serve as the foundational components of each prompt (\autoref{fig:narration}(a1)).
Considering the token limitation of prompts (the maximum number of words or tokens that can be used as input to the model at once), we represent the table in a CSV format.
Moreover, we follow established design guidelines \cite{openai2023} and develop distinct prompts tailored to each specific task.

\subsubsection{Narration with Contextual Information}
Our narration starts with a sentence that describes the chart.
We follow the ``Level 1 semantic content'' framework proposed by Lundgard~\cite{lundgard2021accessible}, which specifies that the description should encompass the chart type, title, legend, encoding channels, and axis information. 
Thus, we design the prompt as illustrated in \autoref{fig:narration}(a2), incorporating the basic chart information from \autoref{fig:narration}(a1) to fulfill our requirements.
The resulting output of this step from \gpt{} will consist of two sentences: the first about the chart type and title and the second addressing the legend, encoding channels, and axis information.
\changed{The example sentences are depicted in \autoref{fig:narration}(a2).}

\subsubsection{Narration with Insights}
Engaging insights is the foundation of a high-quality narration~\cite{srinivasan2019augmenting}.
Following the definitions of facts provided in a prior study~\cite{wang2020DataShot}, we have eight facts as prospective insights for three types of charts: value, proportion, difference, trend, rank, aggregation, extreme, and outlier. 
Within each insight type, there can be various sub-types. 
For example, the proportion insight may encompass majority and minority data. Consequently, we expand the sub-insights, as detailed in \autoref{table:subinsights}, based on the definitions of these eight insights.

\begin{table}[htbp]
\centering
\caption{The table presents the insight type and \\corresponding sub-insight types.}
\begin{tabular}{ll}
\toprule
Insight Type & Sub-insight Type \\ \midrule
Proportion  &  Majority, minority.\\
Extreme   &  Maximum, minimum.\\
Distribution  & Normal, uniform, none.\\
Aggregation & Average, sum, count. \\
Trend & Fluctuate, increase, decrease, \\
& fluctuate increase, fluctuate decrease. \\
\bottomrule
\end{tabular}

\label{table:subinsights}
\end{table}
In the beginning, we expected that \gpt{} would produce several relevant insights from the data table. 
However, this complex task requires a significant amount of input text comprising multiple sentences, which may exceed the maximum allowable number of tokens.
Consequently, we subdivide the task into three sub-tasks: generating all insights, distilling insights, and creating narration.

We first generate all insights with individual prompts for each insight.
These prompts consist of insight definitions (\autoref{fig:narration}(a3)) referenced from previous research~\cite{wang2020DataShot, shi2021autoclips} and basic chart information (\autoref{fig:narration}(a1)).
To ensure efficient data analysis, we require \gpt{} to generate JSON output (\autoref{fig:narration}(b))~\cite{{wei2023leveraging}}. Sometimes, extra explanatory text is added to the output, so we utilize regular expressions to preserve the intended format. Examples can be found in the supplementary.
In this step, we ask \gpt{} to simultaneously identify relevant data for each sub-insight. 
From our experiment and recent researches~\cite{zong2022survey}, \gpt{}'s proficiency in handling mathematical computations, particularly those involving complex calculations (\textit{extreme} and \textit{rank} insights), is limited.
\minor{To address this, we introduce a validation procedure to verify output values. In cases where inaccuracies are identified, heuristic approaches are applied to generate insights, as demonstrated in the supplementary material.}
    
To craft a compelling data narrative, we focus on the salient insights generated by \gpt{}. Thus, we aim to distill insights.
With all insights and the basic chart information (\autoref{fig:narration}(a1)), we task \gpt{} with comparing the relationship between each insight and the visual representation of the data (\autoref{fig:narration}(a4)).
After processing the input data, \gpt{} will generate an output that lists several relevant insight types. 
Taking the example of the ``airport chart'', the result is \textit{``trend''} and \textit{``difference''}.

We then create the narration.
A captivating narrative should interconnect each sentence and flow cohesively rather than being disjointed and independent.
Therefore, we feed all distilled insights into the prompt (\autoref{fig:narration}(a5)), to elicit a coherent and clear narration from \gpt{}.

\subsubsection{Rephrase the Narration}
After acquiring the two segments of narration, we combine them as a new prompt for \gpt{} to refine the final narration (\autoref{fig:narration}(a6)). 
We expect the sentences in the two parts to have smooth transitions between them. 
Keeping the original meaning and essential words, such as the title and axis, \gpt{} will create a refined and cohesive narration by seamlessly connecting the contextual and insightful sentences.
Moreover, we require \gpt{} to re-assign the narration as two parts for the next step. 
\changed{As shown in \autoref{fig:narration}(a6), the red words indicate the adjusted version, and the words in the brackets are omitted.}
Finally, we derive a new narration for contextual information and a new narration for insights.

We then leverage text-to-speech techniques (Azure TTS API\footnote{https://azure.microsoft.com/services/cognitive-services/text-to-speech}) to convert the narration into an audio narrator.

%% file: sections/3.3_animation.tex
\subsection{Animation}
\label{sec:animation}
\begin{figure*}[ht]
    \centering
    \includegraphics{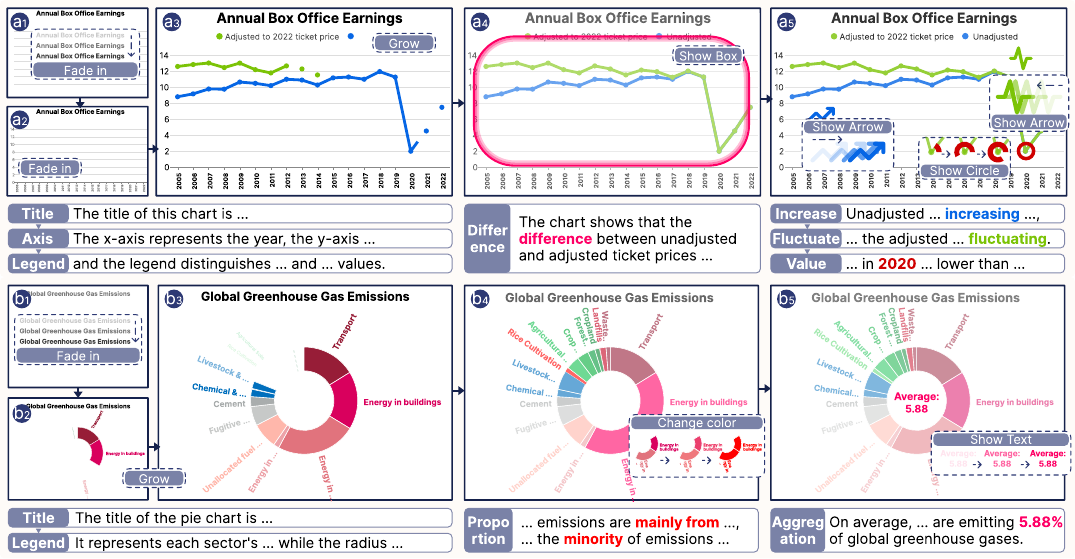}
    \caption{Two use cases. (a1-a5)(b1-b5) The image flow illustrates the keyframes of the Live Chart. Animations are drawn with dotted blue boxes.
    The following text describes the corresponding audio narration, with the first tag indicating the chart component or the insight type.}
    \label{fig:cases}
\end{figure*}

Previous research has demonstrated that animations synchronized with narration can stimulate both visual and auditory senses, leading to improved data representation~\cite{cheng2022investigating, huth2023studies}.
Therefore, we describe how we incorporate animation and synchronize it with narration in this section.
We design the animations based on the two-part narration generated in \autoref{sec:narration}. 
We aim to identify the data related to specific phrases in the narration and add appropriate animations to the corresponding time intervals.
Based on prior research~\cite{ge2020canis}, we use three key aspects to define an animation:
\begin{itemize}
    \item \textbf{Target}: The target elements to be animated.
    As we have added relevant classes and IDs into the input SVG as the labeled SVG (as discussed in \autoref{sec: understanding}), targeting elements is not challenging.
    Technically, we utilize the W3C Selectors API.
    
    \item \textbf{Effects}: The animation effects.
    Similar to \cite{wang2021animated}, we adopt three principal animation types in presentation software (e.g., Keynote and PowerPoint): entrance, emphasis, and exit. 
    Then, we choose multiple prevalent animation styles from past research~\cite{wang2021animated, ge2020canis} and devise styles to fulfill our requirements.
    Part of the animation effects is depicted in \autoref{fig:cases} with dotted blue boxes.
    
    \item \textbf{Interval}: A suitable time interval for elements to enter and exit the screen.
    Instead of using time seconds to represent the entry and exit times, we adopt a different approach. 
    We use the start and end word indices in the narration as a reference.
    For instance, the interval $[10, 20]$ denotes the animation starting from the 10th word and ending at the 20th word.
    Since we use a text-to-speech technique to generate the audio, the API can conveniently generate the timestamp of a specific word in the narration.
    Thus, with all the words having timestamps, we can convert the word index interval into a normal time interval.
    
\end{itemize}

Defining the three variables is the key to completing the task of adding animation.
For each animation, we need to identify the relevant phrases first and convert them into a word index interval. 
Then, we select the corresponding elements as targets and apply an appropriate animation effect to create a polished and visually engaging Live Chart. 
We process the two-part narration separately.

\subsubsection{Animation for the contextual narration}
We first analyze the contextual narration to locate the interval and identify the relevant targets. 
For the first sentence related to the chart title, we set the target as \textit{``title''} and locate the entire first sentence as its interval. 
Moving on to the second sentence about encodings, we locate the position of the words ``x-axis'' and ``y-axis'' for bar and line charts. 
The sentence can be divided into three parts, each representing a distinct interval.
The first and second parts target \textit{``x-axis''} and \textit{``y-axis''}, respectively, and the last part target \textit{``legend''}. 
\changed{For instance, as shown in \autoref{fig:narration}(a6), the first part is represented by ``displays months while'', the second part is represented by ``shows the average... and''.}
For pie charts, we set the interval as the whole sentence and the target as \textit{``sector''} and \textit{``legend''}. 
To display corresponding elements in the front part of live charts, we utilize the entrance animation among the three available animation types.
Thus, animation effects like fade-in and wipe are selected for the contextual narration.

\subsubsection{Animation for the insights}
We create animations that correspond to each insight in order to enhance the storytelling.
Having identified the pertinent data for each sub-insight, we utilized the ``ac-data'' attribute labeled in \autoref{sec: understanding} to target the appropriate elements.
SVG paths and texts can be the candidates.
Concerning the word interval, we continue to rely on \gpt{} to interpret the insightful narration, as it is responsible for generating the narration itself. 
Our approach is progressive, consisting of two prompts. 
Given the table, a list of the distilled insights, and insightful narration, the first prompt is employed to assign sentences to each insight (\autoref{fig:narration}(a7)).
In the output, sentences are included in the insight structure (\autoref{fig:narration}(b)).
Furthermore, with the same inputs (table and insightful narration) and the distilled insights accompanied by their corresponding sentences, the second prompt is utilized to allocate phrases within each insight sentence to each sub-insight (\autoref{fig:narration}(a8)).
\changed{Phrases are added in the subinsights structure after this step (\autoref{fig:narration}(b)).}
Through this process, we identify the appropriate word interval for the animation.
Then we proceed to select the animation effects for the above elements.
We opt for emphasis animations while choosing animation effects. 
We use the animation effects ``show box'', ``show text'', ``change color'' and ``highlight'' for the elements in three chart types.
Additionally, we include the ``bar bounce'' effect for bar charts, as well as ``show arrow'' and ``show circle'' for line and bar charts.
Sometimes certain animation effects may not have specific elements to target within a chart.
For instance, the ``show circle'' effect cannot be used for line charts that lack points.
In such scenarios, we alter the emphasis effect into a list of entrance-exit animations. To be precise, we create such elements and allow them to enter at the beginning of the animation, followed by their exit at the end of the animation.
We select different animation effects for different insights in each chart to ensure the richness of the generated charts.
All animation effects are chosen based on the chart type and insight type.

%% file: sections/4.1_use_cases.tex
\section{Evaluation}
In this section, we first present a set of use cases for reviving real-world charts into \lives{}.
We describe our evaluation process, which involved testing the performance of the understanding model and conducting a user study and expert interview to evaluate \lives{}.
Our evaluation results further confirm the effectiveness of our workflow and demonstrate the potential of the Live Chart format.

\subsection{Use Cases}
\label{sec: usecase}
To demonstrate the capabilities of our workflow in \autoref{sec: livecharts}, we present a set of Live Charts, including bar, line, and pie (donut) charts.
We collected chart SVGs from the Internet and ensured that they covered different topics and styles.
\autoref{fig:teaser} and \autoref{fig:cases} show part of the examples in our gallery~\footnote{The whole examples can be found in the supplementary materials.}.
Each chart is presented in a step-by-step image sequence from index 1 to 5, beginning with contextual information (title, axis, and legend) and followed by various insights.
We use an animation box for each insight depicted in the chart, using a dotted stroke (as shown in \autoref{fig:cases}), to indicate both the type of animation effect and the process by which the corresponding element is altered.
Below the image sequence, we provide the corresponding narration with a tag to indicate contextual or insightful information, followed by a specific description.

%% file: sections/4.2_model_performance.tex
\subsection{Performance of Chart Element Recognition}
\label{sec: performance}

\begin{table*}[ht]
\centering
\caption{The table presents the effectiveness of pixel-based methods and our dual-stream GNN specifically designed for vector graphics recognition on Vega-lite, Plotly, and D3 datasets. $E_e$ denotes only using the element-wise encoder, $E_s$ denotes only using the stroke-wise encoder, and $E_s + E_e$ denotes using the two encoder.}
\begin{tabular}{l|ccc|ccc|ccc}
\toprule
& \multicolumn{3}{c|}{Vega-lite} & \multicolumn{3}{c|}{Plotly} & \multicolumn{3}{c}{D3}\\
        \cmidrule(r){2-4} \cmidrule(l){5-7} \cmidrule(l){8-10}
Methods & $AP_{50}$(\%)  & $AP_{75}$(\%) & mAP (\%) & $AP_{50}$(\%)  & $AP_{75}$(\%) & mAP (\%) & $AP_{50}$(\%)  & $AP_{75}$(\%) & mAP (\%) \\\midrule
YoloV8-m       & 77.30 & 68.90 & 64.60 & 77.60 & 70.60 & \textbf{66.00} & 64.30 & 50.50 & 49.30\\
Ours ($E_e$) & 71.75 & 68.65 & 63.14 & 69.25 & 64.90 & 54.93 & 61.43& 59.77 &  51.98\\
Ours ($E_s$)  & 84.81 & 80.75 & 74.28 & 78.04 & 74.06 & 62.87 & 62.47  & 60.27  & 52.66  \\
Ours ($E_s + E_e$) & \textbf{85.71} & \textbf{81.85} & \textbf{75.21} & \textbf{80.04} & \textbf{76.57} & 64.55 & \textbf{64.77} & \textbf{61.23} & \textbf{53.40} \\
\bottomrule
\end{tabular}

\label{table:AblationStudy}
\end{table*}   

We conducted a series of quantitative evaluations to assess our understanding method in comparison with baselines.

We selected pixel-based methods as our baselines, utilizing mmyolo~\cite{mmyolo2022} for implementation while keeping all parameter settings remaining the same.
Our model is implemented using PyTorch Geometric~\cite{pyg} built upon PyTorch~\cite{PyTorch}. 
We construct a two-layer GNN for the stroke and element-wise encoder. 
We evaluated different machine-learning model parameters and settled on the following sets.
The hidden node representation dimension is set to 64, as deeper GNNs tend to suffer from over-smoothing. 
To avoid over-smoothing, mean aggregation is only conducted on the final layer of the GNN, and each node requires only one transformation and one mean-pooling operation. 
Moreover, a three-layer MLP is employed as the classifier, with the output dimensions of the middle layers set to 512 and 256.
The Adam optimizer is utilized with a learning rate of 0.001 and a batch size of 128. 
The training procedure is performed from scratch for 200 epochs on an Nvidia V100 GPU card.

\textbf{Evaluation Metric.} 
We adopt the commonly used metrics of AP50, AP75, and mAP. Specifically, AP* denotes the average precision at the intersection over the union (IOU) threshold of 0.5 or 0.75 for object detection tasks. Moreover, we calculate the mAP as the mean of the average precision over the IOU thresholds ranging from 0.50 to 0.95.

\textbf{Results.} 
We use the dataset described in \autoref{sec: recognition}.
\minor{For pixel-based methods, we compare our methods with popular one-stage object detection method YoloV8~\cite{yolov8} based on mmyolo.} For YoloV8, the -m variant is scaled YoloV8 with more parameters and better performance. 
As shown in \autoref{table:AblationStudy}, we evaluate the effectiveness of two different vector encoders for chart element recognition. When solely relying on the stroke-wise encoder, our model demonstrates commendable performance. Through our experimentation, we observed that using both stroke-wise and element-wise encoders in conjunction, via the GNN, results in optimal classification performance.

%% file: sections/4.3_user_study.tex
\subsection{User Study}

We further assessed the effectiveness of Live Charts generated by our approach through a user study with 90 participants. 
We conducted our study on Prolific~\footnote{https://www.prolific.co/}, a popular platform for research participation. 
This study aims to assess the following aspects:
\begin{itemize}
    \item Can \lives{} assist users in understanding charts?
    \item In what ways do animation and narration in \lives{} enhance the chart representation?
\end{itemize}
We describe the procedure and results in this section.

\subsubsection{Participants} 
We used the Qualtrics survey platform to set up a crowd-sourcing experiment.
90 participants (29 females) were recruited from Prolific in our study. 
The majority of participants, comprising 42\%, fell into the age range of 25 to 34 years, followed by 27\% in the age range of 35 to 44 years.
As the task required writing, participants' proficiency in writing and English vocabulary could influence the outcomes. Thus, we limited the study to only English-speaking users who had previously received at least a 95\% approval rate on their results. We compensated each participant with \$4.9 for completing the 40-minute task.

\subsubsection{Procedure} 
The study consisted of five parts. 
The first part involved explaining the study procedure and obtaining consent and demographic information from the participants. 
To ensure that our participants had a basic understanding of insights and visualization, we presented eight insight types that we used in our generation process, along with their definitions and examples.
Subsequently, we asked three questions, in which participants needed to observe a chart and answer a question related to three randomly selected insights out of a total of eight.
For example, one of the questions could be, ``What trend best describes the line?''
Only participants who answered these questions correctly were eligible to participate in the study.

To investigate the role of animations and narrations, we carried out a between-subject study. Participants were randomly divided into three groups and presented with varying chart formats: static, animated, and \lives{}.
Each participant was only shown one type of chart format.
For the third part, we demonstrated a bar, pie, and line chart, providing sample insights for each. 
For the fourth part, participants were tasked to provide insights for charts. The chart format used was the same as in the previous step. The six charts are presented randomly.
Participants were asked to fully explore the chart insights and then write down insights.
Moreover, participants with animated charts and Live Charts were required to remain on the page longer than the video length to ensure they viewed it at least once.
Additionally, for participants with Live Charts, we included task questions at random intervals to test the audio functionality, to verify that they can hear the accompanying narration.
we recorded the insights written by users and further calculated the sentence number of insights.

After completing the writing tasks, participants were asked to score the chart. 
They were required to rate the chart in five dimensions: insightful, memorable, focused attention, understandability, and enjoyment.
We used a seven-point Likert scale, with options ranging from strongly disagree (1) to strongly agree (7). 
Finally, participants were asked to provide a brief explanation for their rating.

\subsubsection{Stimuli Preparation}
We collected SVG charts from use cases (\autoref{sec: usecase}), covering a wide range of topics. 
Considering the potential for user fatigue during prolonged experimentation, we prepared a set of 9 charts for each user, with 3 charts for each chart type (bar, pie, and line). Among these, one chart of each type (3 in total) was used for demonstrating examples, and the remaining 6 charts (2 for each type) were used in the formal study.  
For each chart in the study, we prepared three different formats: a static chart, an animated chart, and a \live{}. The original SVG was used as the static version, while the \lives{} were generated using the techniques discussed in \autoref{sec: livecharts}. For the animated charts, we used the same animations as in the \lives{}.
We further shortened the intervals between each animation to one second because sorely removing the narrations from the live version may result in long intervals between animations, which could have bored users.

\subsubsection{Results}
\begin{figure}[htbp]
\includegraphics{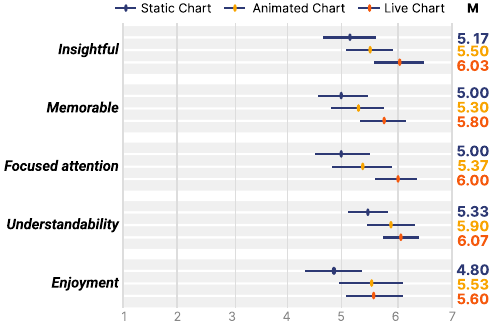}
\centering
\caption{
The statistical analysis results for the overall evaluation include the corresponding Likert ratings. 
The means are depicted using shapes, and their confidence intervals are shown with error bars. On the right side are the specific numerical mean values.}
\label{fig:quatitative}
\end{figure}

\changed{This study gathered the Likert ratings from five perspectives.}
As all data are not normally distributed, we use the Kruskal-Walli test for each factor.
Additionally, we utilized the Mann-Whitney tests with Bonferroni correction to compare pairs within each factor and accounted for multiple comparisons. The sample means, and their corresponding confidence intervals are depicted in \autoref{fig:quatitative} and \autoref{table:quatitative}.

The Likert ratings, measured on a scale from 1 to 7, showed significant differences across all dimensions (p $<$ 0.05). 
Post-hoc testing revealed significantly higher scores for all dimensions in the live format compared to the static charts (p $<$ 0.05, $\left| r \right| > 0.3$). 
Furthermore, we found that Live Charts scored marginally significantly higher than animated charts in the \textit{Insightful} and \textit{Focused Attention} dimensions. Animated charts scored marginally significantly higher than static charts in the \textit{Understandability} and \textit{Enjoyment} dimensions.
The results indicate that Live Charts offer a compelling and memorable approach and effectively direct users' attention toward important information.

\begin{table}[h]
\captionsetup[figure]{labelfont={bf},labelformat={default},labelsep=period,name={Fig.}}
\caption{\minor{Overall statistical evaluation result for each dimension, including the Chi-square values, p-values, degrees of freedom, and effect size. }}
\begin{tabular}{l|l|ll}
\toprule
\multicolumn{1}{c|}{Dimensions}   & Chart Format Pair   & p-value & \thead{effect\\size ($r$)} \\ 
\midrule
\multirow{3}{*}{\thead{Insightful \\ $\chi^2 (2)=8.05$, p=0.017}}      
                                   & Static - Animated & 0.182    & 0.17 \\
                                   & Static - Live     & \textbf{0.008}   & \textbf{0.34}        \\
                                   & Animated - Live   & 0.069    & -0.23       \\
\midrule
\multirow{3}{*}{\thead{Memorable \\ $\chi^2(2)=6.17$, p=0.045}}       
                                   & Static - Animated & 0.284    & 0.13 \\
                                   & Static - Live     & \textbf{0.018}   & \textbf{0.31}        \\
                                   & Animated - Live   & 0.120    & -0.20       \\
\midrule
\multirow{3}{*}{\thead{Focused attention \\ $\chi^2(2)=8.85$, p=0.012}} 
                                   & Static - Animated & 0.206    & 0.16\\
                                   & Static - Live     & \textbf{0.003}   & \textbf{0.38 }       \\
                                   & Animated - Live   & 0.087   & -0.22       \\
\midrule
\multirow{3}{*}{\thead{Understandability \\ $\chi^2(2)=6.36$, p=0.042}} 
                                   & Static - Animated & 0.060    & 0.24\\
                                   & Static - Live     & \textbf{0.017}   & \textbf{0.31}        \\
                                   & Animated - Live   & 0.733    & -0.04       \\
\midrule
\multirow{3}{*}{\thead{Enjoyment \\ $\chi^2(2)=6.18$, p=0.046}}         
                                   & Static - Animated & 0.086    & 0.22\\
                                   & Static - Live     & \textbf{0.013}   & \textbf{0.32}        \\
                                   & Animated - Live   & 0.758    & -0.04      \\
\bottomrule
\end{tabular}
\label{table:quatitative}
\end{table}

\subsubsection{Feedbacks}
Based on our observations, most participants in Live Charts provided positive feedback about Live Charts. 

Both participants in animated charts and Live Charts acknowledged the effectiveness of animation in conveying data effectively.
First, participants using the Live Charts format rated the charts significantly higher in terms of \textit{Understandability} than participants in the static charts format. 
Animated chart users thought that animations allow for breaking down complex information into smaller, more manageable parts, which in return increases comprehension.
Live Chart users also mentioned that \quo{(\lives{}) provide a way to communicate quantitative information in a way that is easier to
 understand for many people.}
In contrast, some participants in static charts found it challenging to absorb all the information presented in a static chart.
For example, one reported that \quo{A chart might be too complicated that I can't recognize all the patterns.}
Moreover, animations provide extra value in the form of accompanying text by extracting information from the static chart using a model. 
This feature was frequently mentioned as a drawback of static charts, with users reporting difficulty in predicting precise numbers.
Second, animated charts and Live Charts users highlighted that animation helps emphasize specific chart elements, transforming data into engaging and memorable stories. 
For example, one participant in Live Charts mentioned that \quo{The animation is also great for highlighting trends and turning the data into a memorable story,} while another participant in static charts wrote that \quo{(Static charts) do not allow highlighting of specific sections.}
Furthermore, users with animated and Live formats agreed that the formats were interesting, while participants with static charts found it dull to look at for a long period. This pattern was also supported in the Likert score for the \textit{Enjoyment} dimension.

Narration emerged as another crucial component, as users noted its ability to complement the animation by directing auditory attention to specific elements.
This pattern can be reflected in the higher \textit{Focused Attention} score in Live Charts.
Moreover, with narration, Live Charts can present specific insights more clearly and effectively, as a participant noted, \quo{(Live Charts) clearly explain what the chart shows.}
Users with animated charts also reinforced this fact by suggesting that \quo{Animated charts can sometimes make it more difficult to understand what is being shown, particularly when only using highlighting to emphasize certain aspects.}
This observation may explain the disparity in the higher score for the \textit{Insightful} dimension for Live Charts.
Furthermore, integrated with the \gpt{} model, the narrator can analyze the data table and combine it with relevant news or context. 
As noted by one user with Live Charts, \quo{There's more context with narration allowing for greater insights into the data and increasing understanding.}

In addition to the benefits of animation and narration on their own, we have also received feedback highlighting the advantages of combining the two.
Users with Live Charts reported that the interplay between animation and narration could enhance the overall experience.
One participant in Live Charts noted that \quo{It is interesting hearing voice while looking at the chart equally.} 
They also suggested that ensuring that the information presented in both the voice-over and the chart matches accurately is crucial.
Furthermore, they rated the \textit{Memorable} dimension significantly higher than participants with the static format, while there was no significant difference between those with animated charts and static charts. 

Although \lives{} offer a more engaging experience by incorporating animations and narrations, they have certain limitations. 
Some users with animated charts and Live Charts commented that \lives{} take up more file space than a static image.
One user noted that \quo{Internet connection and speeds could be a potential drawback,} while another one reported that \quo{(\lives{}) need to have the computer power to display it and cannot just print on paper to display.}
The main reasons for these difficulties can be traced to the characteristics of videos, which consist of a sequence of images.
Additionally, some users with animated charts and Live Charts felt it was not easy to pinpoint the precise moment when a particular insight was discussed, while static charts can be understood at a glance.
This can result in the need to replay the \lives{} multiple times to locate the desired information.
Currently, commercial video players can add small tags to videos' timelines, which can assist users in locating specific content. Such players would be helpful in the playback of Live charts as well.

We also found varying individual preferences for data communication among users.
Although Live Charts was appreciated by a large number of participants for its ability to emphasize particular data points and help absorb the information, some participants pointed out that \lives{} may have limitations in highlighting only certain information, which could be viewed as a drawback, as one noted that \quo{You are more likely to only focus on the information that the narrator is pointing out rather than looking at the chart as a whole for yourself.}
Regarding the method of receiving information, individuals may also exhibit distinct preferences. While some participants enjoy taking an information journey through Live Charts, others prefer a less guided approach to understanding the data by themselves, as noted by one user who stated that \quo{Live Charts don't give people as much time to think on their own.}
The pace of \lives{} can also be divergent. Some participants feel that \quo{(\lives{}) are being explained too slowly.}, while others find that \quo{\lives{} may also go through the information too quickly, and people may miss information.}
Based on the feedback received, we anticipate the development of an authoring tool that would allow users to customize the creation of \lives{} by controlling the information division and the pace for different scenarios.
However, as our research primarily deals with the automatic generation of \lives{}, these concerns are not within the scope of our paper. We will delve deeper into these insightful remarks in the \autoref{sec: discussion_creation}.

%% file: sections/4.4_expert_interview.tex
\subsection{Expert Interview}
We conducted a series of interviews with three expert users from diverse fields to evaluate the effectiveness of our automatic generation workflow. The first expert (E1) has over three years of experience as a data journalist for a digital-news firm. 
\changed{The second expert (E2), possessing a master's degree in visual communication, has over five years of experience. She demonstrates her prowess as a UI/UX designer in a globally acclaimed software firm.}
The third expert (E3) is a senior researcher with over seven years of experience specializing in animation for data visualization.

We used an online meeting system to conduct the interviews, which began with an introduction to Live Charts followed by three examples. We then probed the experts on the quality of Live Charts and their methods for creating such charts. By introducing our automatic workflow, we solicited feedback on its effectiveness. Finally, we asked how it could potentially benefit their work and future endeavors.

All three experts agreed that our Live Charts were of high quality. 
They agreed that the narration logic and overall animation design in Live Charts are appropriate. Specifically, E3 believed that the progressive introduction of information is well-suited for public users, and E2 thought that the animation effects effectively complement the visual design.
E1 stated that \quo{We have previously utilized video formats in our articles, and Live Charts have achieved a similar level of quality to our past efforts.}
The experts also offered suggestions for improving Live Charts, with E1 suggesting that the narration could be more expressive and conversational in tone, rather than sounding too formal and robotic.

Regarding our workflow for automatically generating Live Charts, all the experts appreciated it and found it helpful. They currently use different tools for creating such charts, with E1 adopting existing authoring tools in her company to generate videos, E2 relying on Adobe PR or AE for video editing, and E3 using D3 through coding. 
E1 mentioned that their company's tool generates videos from data, but it may have limitations in terms of chart style. In contrast, our method can handle charts with different styles, including those for online use.  
Additionally, E1 stressed the significance of data accuracy, which our method achieves through the use of models for extraction.
The experts acknowledged the value of our workflow but also offered suggestions for improving it. 
While E1 and E2 suggested incorporating an extension into familiar tools to avoid switching between applications, E1 also raised concerns about data privacy within the company and the risks associated with using external extensions.
Additionally, E1 and E3 recognized the potential for our workflow to transform data news into videos in the future, enabling them to consume data articles more efficiently while performing other tasks.
As an expert in visualization, E3 expressed a preference for a human-AI collaborative approach, in which the user inputs simple insight hints and animation types, and AI assists in making the charts more compact. 

%% file: sections/5_discussion.tex
\section{Discussion}
This section begins with two key lessons learned from the research, specifically related to the format of Live Charts and the generation of Live Charts. 
The limitations of our research are discussed as follows.

\subsection{The Live Chart format}
Reducing cognitive burden is crucial as visualizations become more prevalent across diverse media. Traditionally, individuals relied on their knowledge and experience to understand charts. 
However, the advent of Live Charts can potentially transform this dynamic, offering a new method of information consumption.
According to our evaluation results, \lives{} are enjoyable, tell insightful stories, and improve data understanding.
We received positive feedback from multiple users affirming that Live Charts offer enhanced accessibility for individuals with visual impairments by providing a multi-sensory experience.
Our adopted accessible narration method~\cite{lundgard2021accessible} has proven instrumental in enabling BLVIs to gain a comprehensive understanding of the data presented, even without visual perception.
Moving forward, further advancements can be made by integrating non-speech audio representations of data~\cite{hoque2023accessible, sharif2022voxlens} and efficient navigation strategies~\cite{zong2022richa,thompson2023chart}.
We believe the potential applications of Live Charts are vast, as they can enhance the way information is presented and communicated in various visualizations, including data articles and slideshows.
First, incorporating Live Charts into data articles to replace static ones can transform storytelling, offering readers a more engaging and informative experience. 
Users can access the information conveyed in data news by playing the Live Charts. According to expert feedback, they do not need to be fully focused on the news while watching; instead, they can absorb the information while multitasking.
Slideshows are another example of the application.
As a substitute for static chart images, Live Charts can enliven slides by employing a multi-sensory method of conveying information. For instance, replacing static charts with Live Charts in teaching materials in educational settings can significantly enhance students' learning experiences. This learning process fosters a more profound comprehension of intricate lessons.

\subsection{The creation of Live Charts}
\label{sec: discussion_creation}
We utilize an auto-reviving method that dynamically changes frames with narration guiding users through the information presented.

\textbf{Auto-generated Live Charts}
We have proposed a GNN-based model for recognizing basic charts in SVG format.
Compared with pixel-based charts, vector graphics own higher precision, enabling us to extract data with greater confidence.
Our model shows promise in effectively handling three fundamental chart types compared with pixel-based methods (\autoref{sec: performance}): bar charts, pie charts, and line charts. 
Going forward, we can generalize this model for more chart types~\cite{deng2022kb4va}.
For instance, we anticipate applying this model to scatterplots, which may pose challenges in pixel-based format due to the occlusions. 
Furthermore, existing in-depth analysis~\cite{sarikaya2018scatterplots} underscores an opportunity to recover the embedded data within the scatterplot.
Moreover, while our approach currently shows modest adaptability to infographics with embedded charts~\cite{ying2022glyphcreator, lin2023graph}, it carries the potential for broadening its applicability, encompassing the personalized design of infographics~\cite{ying2022metaglyph}.
We also explored the use of \gpt{}, a popular LLM, to generate text narration for our data visualizations.
For one thing, by adopting a prompt chain containing various insights, the LLM can generate narrations that cater to our needs, deriving fluent and natural-sounding narrations with key insights.
For another, the LLM's training on online news data can assist in inferring the underlying reasons behind certain insights. For instance, in the example of a line chart, as shown in ~\autoref{fig:teaser}(a1-a5), the LLM can infer that ``The box office earnings in 2020 were significantly lower than in previous years due to the COVID-19 pandemic.'' 
This information may not be directly visible in the chart, but with the LLM's assistance, audiences can better understand the context and the story behind the data. Looking ahead, as LLMs continue to evolve and improve, they can potentially incorporate real-time information and help narrations remain relevant and timely. 

\textbf{Opportunities for human-machine collaboration.}
Developing a fully auto-generated Live Chart that satisfies every user's needs and ensures the quality of the narration by GPT-3~\cite{bowman2023eight} is an arduous task, if not impossible, as our user study confirms the challenge due to participants' diverse preferences for Live Chart styles. 
To address this issue, human-machine collaboration is a promising solution~\cite{li2023why, li2023where}. 
We allow users to first select their preferred scenario, tone, and length with the assistant from the LLM. 
Our methods enable the generation of an initial draft of the chart that users can customize to align with their specific narration and animation requirements, reducing the time and effort.
In addition, human involvement becomes crucial for assessing the quality of outcomes, given that LLMs are prone to producing inaccuracies~\cite{ge2023openagi}.
Although existing methods, as discussed in \autoref{sec:narration}, are used to address errors, the rapid evolution of LLMs presents a challenge for automatic error detection as it is not controllable.
Incorporating human judgment into the process maybe a valuable solution, as humans can quickly identify and correct errors at an early stage, avoiding unnecessary downstream efforts. 
Furthermore, there is an expectation for future models to include automated verification features.
We also acknowledge the importance of on-the-fly collaboration, as highlighted by experts we interviewed. To facilitate this, we plan to integrate our functionality as extensions into existing tools, such as data processing applications (e.g., Excel or Tableau) and web browsers. This integration will address the inconvenience of switching between different tools. Users can export charts in diverse formats by incorporating our solution into data processing tools, facilitating data sharing and presentation across various platforms and devices. Meanwhile, integrating with web browsers will enable convenient conversion of online static charts for sharing or presentation purposes.

\subsection{Limitations}
Our research has limitations.
First, we use predefined animations, which may result in limited diversity and somewhat repetitive visual effects.
For example, our highlight animation effect utilized fixed colors without regard to the holistic color scheme. 
In the future, we intend to extend the animation library, ensuring greater diversity and context-aware visual effects.
Second, the generalizability of our understanding model needs improvement. It performs well in dealing with SVGs generated by declarative and imperative language. However, challenges may arise when handling visualizations created with highly flexible tools (e.g., Adobe Illustrator~\footnote{https://www.adobe.com/products/illustrator}) or hand-coded approaches (e.g., Vue, React). Future research could focus on refining the generalizability of our GNN-based approach to address diverse visualization structures.
Third, the outcomes generated by GPT may include errors~\cite{tian2023chartgpt}. 
Despite the use of rule-based methods to minimize errors, these approaches are not exhaustive in resolving all issues~\cite{ge2023openagi}.
Moving forward, we hope that improvements in LLMs will lead to more precise and high-quality answers.
Fourth, we adopt a fixed narration outline in this work for a gradual narration. 
However, we acknowledge the limitations associated with this approach and propose to leave it as future work for LLMs to generate more adaptable outlines.
Fifth, the participants in both the user study and the expert interview have limited coverage.
While we have made efforts by enlisting individuals with varied backgrounds and experiences, we recognize certain limitations, such as the unequal gender distribution in the user study and the absence of specific types of experts in the expert interview. 
In the future, we hope to gain a more comprehensive understanding of how Live Charts support data presentation for a diverse user base in long-term, real-world applications, thereby addressing these limitations.

%% file: sections/conclusion.tex
\section{Conclusion}
In this paper, we have introduced Live Charts as a new format of data presentation that enhances traditional static charts by providing rich animations and audio narration. 
To automate the creation of Live Charts, we decompose complex information from static charts by adopting computer vision techniques and large natural language models. 
Our process first involves extracting data and visual encodings from the chart image, then generating data insights based on this information. We then create narrations and design animations, ultimately combining them to form a Live Chart. 
We conducted various assessments to evaluate the effectiveness of Live Charts and our auto-reviving process, including use cases, chart understanding performance test, a user study, and expert interviews.
The evaluation results demonstrate that Live Charts provide a multi-sensory and engaging experience, with animations and narration significantly contributing to data comprehension. 
Leveraging GPT-4's multimodal input and enhanced performance compared to GPT-3, future research can delve into advanced techniques for automating chart comprehension through LLM and enriching presentations with more vivid narration and animation.